\documentclass[conference]{IEEEtran}
\IEEEoverridecommandlockouts
\usepackage{amsmath}
\usepackage{amsthm}
\usepackage{amssymb}
\usepackage{mathtools}
\usepackage[]{graphicx}    
\usepackage{stmaryrd}
\usepackage{bm}
\usepackage{cite}
\usepackage{amsmath,amssymb,amsfonts}
\usepackage{graphicx}
\usepackage{textcomp}
\usepackage{xcolor}
\usepackage{epsfig}
\usepackage{epstopdf}
\usepackage{balance}
\usepackage{algorithm, tabularx}
\usepackage{float}
\usepackage{algpseudocode}
\usepackage{cuted}
\usepackage{lipsum}
\usepackage{mathtools}
\usepackage{varwidth}
\usepackage[printonlyused]{acronym}
\usepackage{url}
\acrodef{thz}[THz]{Terahertz}
\acrodef{mimo}[MIMO]{multiple-input multiple-output}
\acrodef{los}[LoS]{line-of-sight}
\acrodef{nlos}[NLoS]{non-line-of-sight}
\acrodef{mimo}[MIMO]{multiple-input multiple-output}
\acrodef{ofdm}[OFDM]{orthogonal frequency-division multiplexing}
\acrodef{tx}[Tx]{transmitter}
\acrodef{rx}[Rx]{receiver}
\acrodef{aoa}[AoA]{angle of arrival}
\acrodef{aod}[AoD]{angle of departure}
\acrodef{ula}[ULA]{uniform linear array}

\DeclareMathOperator*{\argmin}{arg\,min}
\newcommand{\ignore}[1]{}  
\newcommand{\RR}{\mathbb{R}}
\newcommand{\CC}{\mathbb{C}}
\newcommand{\NN}{\mathbb{N}}
\newcommand{\XX}{\bm{\mathcal{X}}}
\newcommand{\YY}{\bm{\mathcal{Y}}}

\def\BibTeX{{\rm B\kern-.05em{\sc i\kern-.025em b}\kern-.08em
    T\kern-.1667em\lower.7ex\hbox{E}\kern-.125emX}}
\begin{document}

\title{Tensor-based Space Debris Detection for \\ Satellite Mega-constellations}

\author{\IEEEauthorblockN{Olivier Daoust\IEEEauthorrefmark{1} Hasan Nayir\IEEEauthorrefmark{2}, Irfan Azam\IEEEauthorrefmark{1}, Antoine Lesage-Landry\IEEEauthorrefmark{3}, Gunes Karabulut Kurt\IEEEauthorrefmark{1}}
\IEEEauthorblockA{\IEEEauthorrefmark{1}Poly-Grames Research Center, Department of Electrical Engineering, Polytechnique Montréal, Montréal, QC, Canada}
\IEEEauthorblockA{\IEEEauthorrefmark{2}Department of Electronics and Communication Engineering, Istanbul Technical University, {\.{I}}stanbul, Turkey}
\IEEEauthorblockA{\IEEEauthorrefmark{3}Department of Electrical Engineering, Polytechnique Montréal, MILA \& GERAD, Montréal, QC, Canada}
Emails:\texttt{\{olivier.daoust,irfan.azam,antoine.lesage-landry,gunes.kurt\}@polymtl.ca,} \\ \texttt{nayir20@itu.edu.tr}}
\maketitle

\begin{abstract}
Thousands of satellites, asteroids, and rocket bodies break, collide, or degrade, resulting in large amounts of space debris in low Earth orbit. The presence of space debris poses a serious threat to satellite mega-constellations and to future space missions. Debris can be avoided if detected within the safety range of a satellite. In this paper, an integrated sensing and communication technique is proposed to detect space debris for satellite mega-constellations. The canonical polyadic (CP) tensor decomposition method is used to estimate the rank of the tensor that denotes the number of paths including line-of-sight and non-line-of-sight by exploiting the sparsity of THz channel with limited scattering. The analysis reveals that the reflected signals of the THz can be utilized for the detection of space debris. The CP decomposition is cast as an optimization problem and solved using the alternating least square (ALS) algorithm. Simulation results show that the probability of detection of the proposed tensor-based scheme is higher than the conventional energy-based detection scheme for the space debris detection.
\end{abstract}


\section{Introduction}
A large number of satellites in the low Earth orbit (LEO) have been planned by the space industries to create satellite mega-constellations. Many of these space missions may unfortunately lead to a significant amount of space debris, including the abandoned parts of rockets or broken fragments of satellites due to collisions. Additionally, some of the naturally occurring celestial bodies such as asteroids are contributing to the congestion around Earth \cite{nature2021,deepspace2023}. It has been estimated that there are approximately 100 million space debris larger than 1~mm orbiting the Earth. Because the orbiting speed in LEO is around 7.8 km/s, very small debris can cause great damage to the payload \cite{NASA2023}. Hence, space debris constitute an important threat to the orbiting space stations and satellites, and to future launches. Therefore, efficient space debris detection techniques are required to prevent collisions and to preserve these high-cost space equipment. 
 
Integrated sensing and communication (ISAC) techniques are now at the forefront of development of the future integrated radar sensing and wireless communication systems~\cite{hanzo2020,isac2022}. Importantly, the use of millimeter wave (mmWave) and terahertz (THz) frequencies in these systems presents significant challenges due to the high level of signal attenuation for this frequency range \cite{shafie2022terahertz}. To overcome this, large antenna arrays are used at both the transmitters and at the receivers under the massive multiple input multiple output (MIMO) system paradigm. This allows beamforming, which focuses the transmission and reception of the signal in specific directions, to be used to provide enough gain to compensate for the path loss hence maintaining a strong signal \cite{akyildiz2018combating}. These challenges become even more prominent in the context of line-of-sight (LOS) and non-line-of-sight (NLOS) THz links \cite{moldovan2014and}. Because NLOS links have poor propagation characteristics, they are better suited to sensing than to communication in ISAC. From a reflection standpoint, in NLOS, it is usually assumed that the signal is perpendicular to the plane of incidence in terms of the transverse electric part of the electromagnetic wave \cite{jcas_icc2023}. However, the angle at which the THz signal encounters space debris can affect the nature of the reflection. This angle can provide valuable insights such as the size, the composition, and the trajectory of the space debris. In order to develop more accurate and reliable detection techniques, the characteristics of signal reflection, including the angle of incidence should be considered. The selection of a suitable waveform is also critical in this context \cite{Gizem_GB2022}. Hence, understanding the angle of reflection along with the appropriate waveform is essential for accurately interpreting the reflected signals for space debris detection.

Recently, tensor-based techniques have been extensively investigated for signal processing. Compared to the matrix-based methods, tensors can extract more significant data components from multidimensional data \cite{tensor2017,tensorML}. Namely, canonical polyadic (CP) decomposition is mostly utilized for channel and target parameter estimations in MIMO radar systems \cite{tensorMIMO}. Moreover, tensors exhibit excellent performance when introduced into ISAC \cite{jcas_icc2023}. Specifically, the environment sensing parameters can be extracted through the rank of tensors by leveraging the sparse nature of the THz channel when used in ISAC systems. However, rank detection in THz massive MIMO systems can be challenging due to hybrid precoding structures and to the large number of antennas. Recently, nested tensor-based technique is introduced in the reconfigurable intelligent surface empowered ISAC system that is able to perform the symbol detection as well as target localization to avoid the cumbersome pilot transmission process \cite{isac_ris2023}.  Additionally, tensor decomposition has been found effective for the estimation of the channel parameters and the location of moving targets \cite{tensor_iot}. Many of the previous research works still lack the investigations on the effectiveness of tensors in ISAC considering the challenges of the inter-satellites links particularly, the space debris detection. Traditional methods of space debris detection often rely on radar systems and have limited coverage especially for objects in higher orbits. Also, the ground-based telescopes are inefficient due to the limited visibility and are less accurate. Therefore, in this work, we propose a tensor-based ISAC system for satellites in mega-constellations using the THz band that provides better accuracy and resolution for space debris detection.


\subsection{Motivation and Contribution} 
In this paper, a tensor decomposition method known as the canonical polyadic decomposition (CPD) is used for detecting space debris for satellite mega-constellations. Particularly, it relies on the following three key observations: 

\begin{enumerate}
	\item By having a straightforward setup for LEO satellite transmission, the received signal can be structured into a tensor of third order, which admits a CP decomposition.
	\item The tensor has a low CP rank because of the sparse nature of THz and space channels, making the CP decomposition possible. The rank of the constructed tensor represents the total number of paths including the direct LOS and the NLOS paths reflected by the space debris. 
	\item The rank of the constructed tensor truly depends on the channel parameters, such as angle of arrival (AoA), angle of departure (AoD), time delay, and fading coefficients. Thanks to use of tensors for data representation and processing that leads to a very low computational complexity.
\end{enumerate}
 
\begin{figure*}
	\centering
	\includegraphics[width=0.99\linewidth]{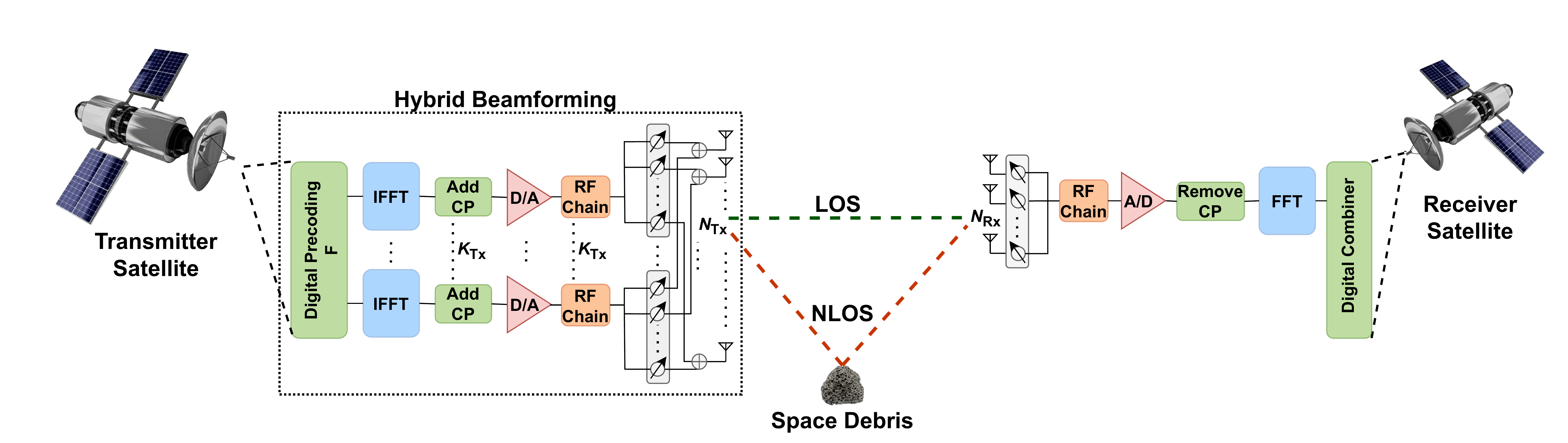}
	\caption{The system architecture of the massive MIMO-OFDM enabled satellite mega-constellations operating at THz frequencies for space debris detection.}
	\label{fig:main_figure}
\end{figure*}
 
The main contribution of this paper is the space debris detection using CP decomposition for LEO satellites. In the proposed system, the THz channel is investigated and its multipath characteristic is utilized to detect the space debris. Importantly, the tensor model and the problem formulation for space debris detection are developed considering the channel conditions for LEO satellites. Moreover, the alternating least square (ALS) algorithm is presented for CP decomposition to estimate the rank of the tensor. A tensor-based numerical analysis followed by the Monte Carlo simulation is performed to estimate the probability of detection and the performance of the proposed system is compared with a conventional detection technique. The results illustrate the effectiveness and superiority of the proposed tensor-based space debris detection in comparison to the conventional energy-based detection.

The rest of the paper is organized as follows. Section~2 presents the system model of our proposed space debris detection system for LEO satellites. Section~3 formulates the tensor-based detection scheme, including the CP decomposition and the optimization problem used to perform the decomposition. Performance evaluation and simulation results are presented in Section~4. Finally, Section~5, draws conclusions and outlines future directions.

\section{System Model}
This study considers a THz massive MIMO-OFDM system consisting of two satellites, one of which is a receiver, and the other is a transmitter. Each satellite employs a hybrid beamforming structure that combines analog and digital components to facilitate hardware implementation. In this case, we assume that the \ac{tx} satellite has $N_{\text{Tx}} \in \mathbb{N}$ antennas and $K_{\text{Tx}} \in \mathbb{N}$ RF chains, and that the \ac{rx} satellite has $N_{\text{Rx}} \in \mathbb{N}$ antennas and $K_{\text{Rx}} \in \mathbb{N}$ RF chains. Specifically the \ac{rx} satellite has only one RF chain. The number of OFDM subcarriers $K \in \mathbb{N}$ are selected for training from the total number of subcarriers $\bar{K} \in \mathbb{N}$. Moreover, the \ac{tx} satellite uses the $T \in \mathbb{N}$ different beamforming vectors at $T$ successive time frames. Also, each time frame is divided into $M \in \mathbb{N}$ sub-frames, and the \ac{rx} satellite uses and individual combining vector at each sub-frame. We can represent the beamforming vector $\mathbf{x}_k(t) \in \CC^{K_{\text{Tx}}\times N_{\text{Tx}}}$ related with the \emph{k}$^{\text{th}}$ subcarrier at the \emph{t}$^{\text{th}}$ time frame as:
\begin{equation}
\mathbf{x}_k(t)=\mathbf{F}_{\text{RF}}(t) \mathbf{F}_k(t) \mathbf{s}_k(t), \quad \forall k=\{1, \ldots, K\},
\end{equation}
where $\mathbf{s}_k(t) \in \CC^{K}$ denotes the pilot symbol vector, $\mathbf{F}_k(t) \in \CC^{K_{\text{Tx}} \times K}$ denotes the digital precoding matrix for the \emph{k}$^{\text{th}}$ subcarrier, and $\mathbf{F}_{\text{RF}}(t) \in \CC^{N_{\text{Tx}} \times K_{\text{Tx}}}$ is the RF precoder for all subcarriers. In order to generate the beamforming vector (1), at first, the pilot symbol $\mathbf{s}_k(t)$ is precoded using digital precoding matrix $\mathbf{F}_k(t)$, Then, the symbol blocks using inverse discrete Fourier transform (IDFT) are transformed to the time-domain followed by the addition of a cyclic prefix. All the subcarriers are precoded with the RF precoder $\mathbf{F}_{\text{RF}}(t)$ at the end.  

The \ac{rx} satellite employs $M$ RF combining vectors $\mathbf{q}_{m} \in \CC^{N_{\text{Rx}}}$ to detect the transmitted signal at each time frame. The received signal is combined in the RF-domain. The symbols are converted back to frequency-domain, separating the different subcarriers using a discrete Fourier transform (DFT) after the removal of the cyclic prefix. The received signal $y_{k, m}(t)$ associated with the \emph{k}$^{\text{th}}$ subcarrier at the \emph{m}$^{\text{th}}$ sub-frame is expressed as:
\begin{equation}
y_{k, m}(t)=\mathbf{q}_m^\top \mathbf{H}_k \mathbf{x}_k(t)+w_{k, m}(t),
\end{equation}
where $\mathbf{q}_{m} \in \CC^{N_{Rx}}$ identifies the combining vector used at the \emph{m}$^{\text{th}}$ sub-frame, $\mathbf{H}_{k} \in \CC^{N_{Rx} \times N_{Tx}}$ is the channel matrix associated with the $k^{\text{th}}$ subcarrier, and $w_{k, m}(t)$ denotes the additive Gaussian noise. Collecting the $M$ received signals $\left\{y_{k, m}(t)\right\}_{m=1}^M$ at each time frame, at the vector $\mathbf{y}_k(t)$, we have
\begin{equation}
\mathbf{y}_k(t)=\mathbf{Q}^\top \mathbf{H}_k \mathbf{x}_k(t)+\mathbf{w}_k(t),
\end{equation}
where 
\begin{equation}
\begin{aligned}
\mathbf{y}_k(t) & \triangleq\left[\begin{array}{lll}
y_{k, 1}(t) & \ldots & y_{k, M}(t)
\end{array}\right]^\top, \\
\mathbf{w}_k(t) & \triangleq\left[\begin{array}{llll}
w_{k, 1}(t) & \ldots & w_{k, M}(t)
\end{array}\right]^\top, \\
\mathbf{Q} & \triangleq\left[\begin{array}{lll}
q_1 & \ldots & q_M
\end{array}\right].
\end{aligned}
\end{equation}

\subsection{Channel Model}
It has been investigated that THz channels in  space environment typically exhibit limited scattering characteristic \cite{Thzspace}. Therefore, we adopt a geometric THz channel model with $L  \in \mathbb{N}$ scatterers between two satellites. Each path component is characterized by a time delay, AoA, and AoD, $\tau_{l} \in \RR,\theta_l, \phi_l \in[0,2 \pi]$, respectively. Using these parameters, the channel matrix in the delay domain is expressed as:
\begin{equation}
\mathbf{H}(\tau)=\sum_{l=1}^L \bm{\alpha}_l \mathbf{a}_{\text{MS}}\left(\theta_l\right) \mathbf{a}_{\text{BS}}^\top \left(\phi_l\right) \delta\left(\tau-\tau_l\right),
\end{equation}
where $\bm{\alpha}_l \in \CC^L $ is the complex path gain, $\delta$($\cdot$) is the delta function,  $\mathbf{a}_{\text{MS}}\left(\theta_l\right)$, $\mathbf{a}_{\text{BS}}^\top \left(\phi_l\right)$, denote the antenna array response vectors of the \ac{tx} and \ac{rx} satellites, respectively. Finally, the frequency domain channel matrix $\mathbf{H}_{k}$ associated with the \emph{k}$^{\text{th}}$ subcarrier is represented by
\begin{equation}\label{channel_frequency}
\mathbf{H}_k=\sum_{l=1}^L \bm{\alpha}_l \mathrm{e}\!\left(-j 2 \pi \tau_l f_\text{s} K / \bar{K}\right) \mathbf{a}_{\text{MS}}\left(\theta_l\right) \mathbf{a}_{\text{BS}}^\top \left(\phi_l\right),
\end{equation}
where $f_\text{s}$ is the sampling rate.

\subsection{Path Loss Model}
The signal can propagate as LOS or as NLOS in the communication channels. In the case of an LOS transmission, the channel gain is \cite{chaccour2022can}:
\begin{equation}
\bm{\alpha}_{n, u}^L=\frac{c}{4 \pi f r_{n, u}} \mathrm{e}^{-\frac{k(f) r_{n, u}}{2}} \mathrm{e}^{-j 2 \pi f \tau_{n, u}^L},
\end{equation}
where \emph{c} is the speed of the light, $k(f) \in \mathbb{R}$ is the molecular absorption coefficients of the medium, $f > 0$ is the operating frequency, and $r_{n,u} > 0$ is the distance between the \ac{tx} satellite and \ac{rx} satellite. We consider limited NLOS links due to the poor propagation characteristics of THz signals. In NLOS conditions, the channel gain is expressed as \cite{moldovan2014and}:
\begin{equation}
\bm{\alpha}_{n, u}^N\!=\!\frac{c}{4 \pi f\left(r_{n, u}^{(1)}+r_{n, u}^{(2)}\right)} \mathrm{e}^{\!\left(\!\!-\frac{k(f)\left(r_{n, u}^{(1)}+r_{n,u}^{(2)}\right)}{2}\!\right)}\! R(f) \mathrm{e}^{-j 2 \pi f \tau_{n, u}^N},
\end{equation}
where $r_{n, u}^{(1)} > 0$ and $r_{n, u}^{(2)} > 0$ is the distance between the \ac{tx} satellite and the space debris, and the distance between the space debris and the \ac{rx} satellite, respectively. In addition, $R_{n, u}(f)=\gamma_{n, u}(f) \rho_{n, u}(f)$ is the reflection coefficient, where $\gamma(f) \approx-\mathrm{e} \left(\frac{-2 \cos \left(\psi_{n, u}\right)}{\sqrt{\eta(f)^2-1}}\right)$ is the Fresnel reflection coefficient and $\rho_{n, u}(f)=\mathrm{e} \left(-\frac{8 \pi^2 f^2 \sigma^2 \cos ^2\left(\psi_{n, u}\right)}{c^2}\right)$ is the Rayleigh factor which characterizes the roughness effect. Finally, $\psi_{n, u}$ is the angle of the signal coming towards to the reflector, $\eta(f)$ is the refractive index, and $\sigma$ is the surface height standard deviation.

\section{Tensor-based Detection}
In this section, we introduce our proposed tensor-based space debris detection approach. It includes a tensor decomposition method and the ALS algorithm for CP decomposition to estimate the rank of the tensor.

\subsection{Tensors and the Canonical Polyadic Decomposition}
Vectors and matrices can be considered as first-order and second-order tensors, respectively. Arrays of higher order are usually referred to as tensors. In this paper, we mainly focus on low-rank third-order tensors. A tensor defined as $\XX$ $\in$ $\CC^{I \times J \times K}$, $I, J, K \in \NN, $ where each entry is denoted as  $\XX(i,~j,~k)$ for $i=\{1,2,\dots,I\}$, $j=\{1,2,\dots,J\}$, and $k=\{1,2,\dots ,K$\}. A rank-1 third-order tensor is expressed as:
\begin{equation}
    \bm{\XX} = \mathbf{a} \circ \mathbf{b} \circ \mathbf{c} ,
\end{equation}
where ${\mathbf{a}} \in {\CC}^{I}$, $\mathbf{b} \in \CC$$^{J}$, $\mathbf{c} \in \CC$$^{K}$, and $\circ$ denotes the outer product.
The rank of a tensor $\bm{\XX}$ represents the number of rank-1 tensors needed to construct $\bm{\XX}$ as their sum~\cite{tensorML}. A tensor $\bm{\XX}$ $\in$ $\CC$ $^{I \times J \times K}$ can be represented as the sum of the outer product between three vectors:
\begin{equation}
    \bm{\XX}=\sum_{\textit{f}=1}^{\textit{F}} {\mathbf{a}}_f \circ {\mathbf{b}}_f \circ {\mathbf{c}}_f,
\end{equation}
where $F \in \NN$ denotes the rank of $\XX$. Let ${\mathbf{A}} \in {\CC}^{I\times F}$, $\mathbf{B} \in \CC$$^{J \times F}$, $\mathbf{C} \in \CC$$^{K\times F}$ where $\mathbf{A}=[\mathbf{a}_1,\mathbf{a}_2,\dots,\mathbf{a}_F]$, $\mathbf{B}=[\mathbf{b}_1,\mathbf{b}_2,\dots,\mathbf{b}_F]$, and $\mathbf{C}=[\mathbf{c}_1,\mathbf{c}_2,\dots,\mathbf{c}_F]$. We can rewrite (10) as:
\begin{equation}
    \bm{\XX}(i,j,k)=\sum_{\textit{f}=1}^{\textit{F}} \textbf{A}(i,f)~\textbf{B}(j,f)~\textbf{C}(k,f).
\end{equation}
For conciseness, (11) can be expressed as $\bm{\XX}$ = $\llbracket \textbf{A},~\textbf{B},~\textbf{C} \rrbracket$. The process of decomposing a tensor into a set of matrices is referred to as the \textit{polyadic decomposition}. The decomposition used in (11) is referred to as the CPD. In this work, we leverage the CPD for space debris detection.

\subsection{Rank Estimation of Tensor with CPD}
We can express the \emph{k}$^{\text{th}}$ subcarrier received signal as
\begin{equation}
\mathbf{Y}_k=\mathbf{Q}^\top \mathbf{H}_k \mathbf{P}+\mathbf{W}_k, \quad k=\{1, \ldots, K\}
\end{equation}
where
\begin{equation}
\begin{array}{l}
\mathbf{Y_k} \triangleq\left[\mathbf{y}_k(1) \quad \ldots \quad \mathbf{y}_k(T)\right],\\
\mathbf{W_k} \triangleq\left[\mathbf{w}_k(1) \quad \ldots \quad \mathbf{w}_k(T)\right],\\
\mathbf{P} \triangleq[\mathbf{p}(1) \quad \ldots \quad \mathbf{p}(T)].
\end{array}
\end{equation}
We can represent the received signal as third-order tensor $\YY_k \in \CC^{M \times T \times K}$. Using (6) and (12), we obtain
\begin{equation}
\YY_k=\sum_{l=1}^L \tilde{\bm{\alpha}}_{l, k} \tilde{\mathbf{a}}_{\text{MS}}\left(\theta_l\right) \tilde{\mathbf{a}}_{\text{BS}}^\top \left(\phi_l\right)+\mathbf{W}_k,
\end{equation}
where $\tilde{\bm{\alpha}}_{l, k} \triangleq \bm{\alpha}_l \mathrm{e} \left(-j 2 \pi \tau_l f_\text{s} k / \bar{K}\right)$, $\tilde{\mathbf{a}}_{\text{MS}}\left(\theta_l\right) \triangleq \mathbf{Q}^\top \mathbf{a}_{\text{MS}}\left(\theta_l\right)$, and $\tilde{\mathbf{a}}_{\text{BS}}\left(\phi_l\right) \triangleq \mathbf{P}^\top \mathbf{a}_{\text{BS}}\left(\phi_l\right)$.
Using CP decomposition which decomposes a tensor into a sum of rank-one component tensors, the tensor $\YY$ is expressed as:
\begin{equation}
{\YY}=\sum_{l=1}^L \tilde{\mathbf{a}}_{\mathrm{MS}}\left(\theta_l\right) \circ \tilde{\mathbf{a}}_{\mathrm{BS}}\left(\phi_l\right) \circ \left(\bm{\alpha}_l g\left(\tau_l\right)\right)+\mathbf{W},
\end{equation}
where,
\begin{equation}
g \! \left(\tau_l\right)\!\triangleq\!\left[\mathrm{e} \left(-j 2 \pi \tau_l f_\text{s}(1 / \bar{K})\right) \ldots \mathrm{e} \left(-j 2 \pi \tau_l f_\text{s}(K / \bar{K})\right)\right]^\top \!.
\end{equation}
Hence, decomposing $\YY$ and then computing the rank on the resulting matrices leads to the $L$ number of paths. Consequently, if $L > 1$, a reflection on a space debris has occurred.

We are thus interested in computing $L$. Let $\hat{L}>L$ be an overestimated guess for the rank of $\YY$. We decompose $\YY$ into its rank-$\hat{L}$ canonical polyadic form and use a Frobenius-norm regularizer to promote low-rank solutions. If $L<\hat{L}$, a well-tuned regularized decomposition should identify the rank-$L$ ${\mathbf{A}} \in {\CC}^{I\times F}$, $\mathbf{B} \in \CC$$^{J \times F}$, and $\mathbf{C} \in \CC$$^{K\times F}$ matrices. The CPD of $\YY$ can be computed by the following trilinear optimization problem~\cite{tensor2017}:
\begin{align}
    &{{\min_{\mathbf{\hat{\mathbf{A}},\hat{\mathbf{B}},\hat{\mathbf{C}}}}} \left\|{\bm{\YY}} - \bm{\XX} \right\|_F^2 + \mu \left( \left\| \hat{\mathbf{A}} \right\|_F^2 +  \left\| \hat{\mathbf{B}} \right\|_F^2 +  \left\| \hat{\mathbf{C}} \right\|_F^2 \right)} \nonumber\\
    &\text{subject to}~\bm{\XX} = \sum_{l=1}^{\hat{L}} \mathbf{\hat{a}}_l~\circ~\mathbf{\hat{b}}_l~\circ~\mathbf{\hat{c}}_l,\\
    & \qquad\qquad~~ \mathbf{\hat{\mathbf{A}}}=[\mathbf{\hat{a}}_1, \mathbf{\hat{a}}_2, \text{\dots}, \mathbf{\hat{a}}_{\hat{L}}], \nonumber\\
    & \qquad\qquad~~\mathbf{\hat{\mathbf{B}}}=[\mathbf{\hat{b}}_1, \mathbf{\hat{b}}_2,\text{\dots}, \mathbf{\hat{b}}_{\hat{L}}], \nonumber\\
    & \qquad\qquad~~ \mathbf{\hat{\mathbf{C}}}=[\mathbf{\hat{c}}_1, \mathbf{\hat{c}}_2,\text{\dots}, \mathbf{\hat{c}}_{\hat{L}}]. \nonumber
\end{align}
where $\left\| \cdot \right\|_F$ denotes the Frobenius-norm and $\mu > 0$ is a regularization parameter. The optimization problem (17) can be solved using an alternating least square (ALS) method~\cite{tensor2017}. The step by step procedure is detailed in the Algorithm~\ref{alg:cpd_als}. 

In Algorithm~\ref{alg:cpd_als}, $\epsilon  > 0$ is the maximum tolerance factor used as the convergence criterion to determine when the Algorithm~\ref{alg:cpd_als} reaches to acceptable values of $\mathbf{\hat{\mathbf{A}}}$, $\mathbf{\hat{\mathbf{B}}}$, and $\mathbf{\hat{\mathbf{C}}}$, $T_{max} \in \NN$ is the maximum number of iterations, the symbol $\odot$ is the Khatri-Rao product, and the function \texttt{cpdgevd} is a subroutine which uses the generalized eigenvalue decomposition to provide a good initialization value for solving the non-convex problem (17) using  ALS~\cite{tensorlab3.0}. Algorithm~\ref{alg:cpd_als} returns a set $U_{\hat{L}}$ which contains $\mathbf{\hat{\mathbf{A}}^*}$, $\mathbf{\hat{\mathbf{B}}^*}$, and $\mathbf{\hat{\mathbf{C}}^*}$. After the decomposition, if $L < \hat{L}$, then $\mathbf{\hat{\mathbf{A}}^*}$, $\mathbf{\hat{\mathbf{B}}^*}$, and $\mathbf{\hat{\mathbf{C}}^*}$ should not have full column rank. Finally, we extract full column rank from the $\mathbf{\hat{\mathbf{A}}^*}$, $\mathbf{\hat{\mathbf{B}}^*}$, and $\mathbf{\hat{\mathbf{C}}^*}$ matrices by thresholding their associated singular values. Computing the rank of the resulting matrices then leads to an estimation of $\YY$'s rank and thus, the number of reflected paths.

\begin{algorithm}[t!]
	\caption{Regularized ALS for tensor rank estimation}
	\begin{algorithmic}
		\State  $\textbf{Parameters:}~I,~J,~K,~\epsilon,~T_{max}$
		\State  $\textbf{Initialization:}~\text{Set}~ it=0$
		\For {$i=1:I$} 
		\State $\mathbf{Y}_1 = \texttt{vec}(\bm{\YY}(i,:,:))$
		\EndFor
		\For {$j=1:J$} 
		\State $\mathbf{Y}_2 = \texttt{vec}(\bm{\YY}(:,j,:))$
		\EndFor
		\For {$k=1:K$} 
		\State $\mathbf{Y}_3 = \texttt{vec}(\bm{\YY}(:,:,k))$
		\EndFor
		\State $U_0 = \texttt{cpdgevd}(\bm{\YY},\hat{L})$
		\State $[\hat{\mathbf{A}},\hat{\mathbf{B}},\hat{\mathbf{C}}]=[U_0\{1\},U_0\{2\},U_0\{3\}]$\\
		\While{$\left\| \llbracket \hat{\mathbf{A}},\hat{\mathbf{B}},\hat{\mathbf{C}} \rrbracket - \bm{\YY} \right\|_F \geq \epsilon$ \& $it \leq T_{max}$}
		\State 
		\begin{align*}
			\hat{\mathbf{A}} &\leftarrow \argmin_{{\mathbf{A}} \in {\CC}^{I\times F}} \left\| \mathbf{Y}_1 - (\hat{\mathbf{C}} \odot \hat{\mathbf{B}})\mathbf{A}^\top \right\| ^2 _F + \mu \lVert \mathbf{A} \rVert ^2 _F \\
			\hat{\mathbf{B}} &\leftarrow \argmin_{{\mathbf{B}} \in {\CC}^{J\times F}} \left\| \mathbf{Y}_2 - (\hat{\mathbf{C}} \odot \hat{\mathbf{A}})\mathbf{B}^\top \right\| ^2 _F + \mu \left\| \mathbf{B} \right\| ^2 _F \\
			\hat{\mathbf{C}} &\leftarrow{} \argmin_{{\mathbf{C}} \in {\CC}^{K\times F}} \left\| \mathbf{Y}_3 - (\hat{\mathbf{B}} \odot \hat{\mathbf{A}}){\mathbf{C}}^\top \right\| ^2 _F + \mu \lVert \mathbf{C} \rVert ^2 _F\\  
			it &\leftarrow it+1
		\end{align*}
		\EndWhile
		\State $\textbf{Return}~U_{\hat{L}}$
	\end{algorithmic}
	\label{alg:cpd_als}
\end{algorithm}

\section{Simulation Results}
In this section, the performance of the proposed tensor-based detection approach (referred to as TBD) is evaluated and compared with the conventional energy-based detection scheme (referred to as EBD). We consider a scenario where the \ac{tx} satellite employs a \ac{ula} with 64 antennas and the \ac{rx} satellite employs a \ac{ula} with 32 antennas. The distance between antenna elements is set to be half of the wavelength. In our simulation, the THz channel is generated using the ray-tracing tool, according to the \ac{tx}-\ac{rx} satellite pairs placed at different points. The AoA ($\theta_l$), AoD ($\phi_l$), number of path ($L$), and the delay spread ($\tau_l$) are obtained from the ray-tracing tool \cite{raytracing}, and are used in (\ref{channel_frequency}). In addition, since THz band communication systems are more dependent on temperature changes, we let the noise model be a function of the solar brightness temperature in the deep space. Thus, the noise temperature is the sum of the brightness temperature and the thermal noise temperature \cite{nie2021channel} 
\begin{equation}
T_N=T_b+T_0,
\end{equation}
where $T_b$ and $T_0$ are the solar brightness temperature and the ambient noise temperature, respectively. The number of subcarriers $K=6$ are selected for training from the total number of subcarriers $\bar{K}=128$. Moreover, the beamforming matrix $\mathbf{P}$ and the combining matrix $\mathbf{Q}$ are randomly generated using uniformly selected entries from a unit circle. Main simulation parameters are listed in Table \ref{table1}.

\begin{figure}[t!]
	\includegraphics[scale=0.6]{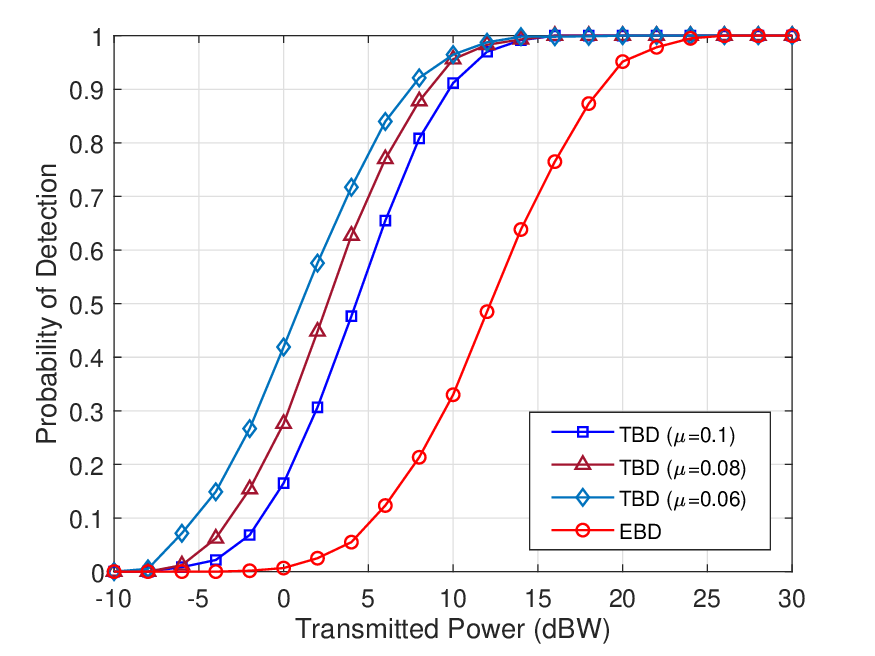}
	\caption{Probability of detection of the proposed TBD with different regularization values $\mu$=$[0.06,0.08,0.1]$ and the conventional EBD scheme.}
	\label{result:fig1}
\end{figure}

\begin{figure}[t!]
	\includegraphics[scale=0.6]{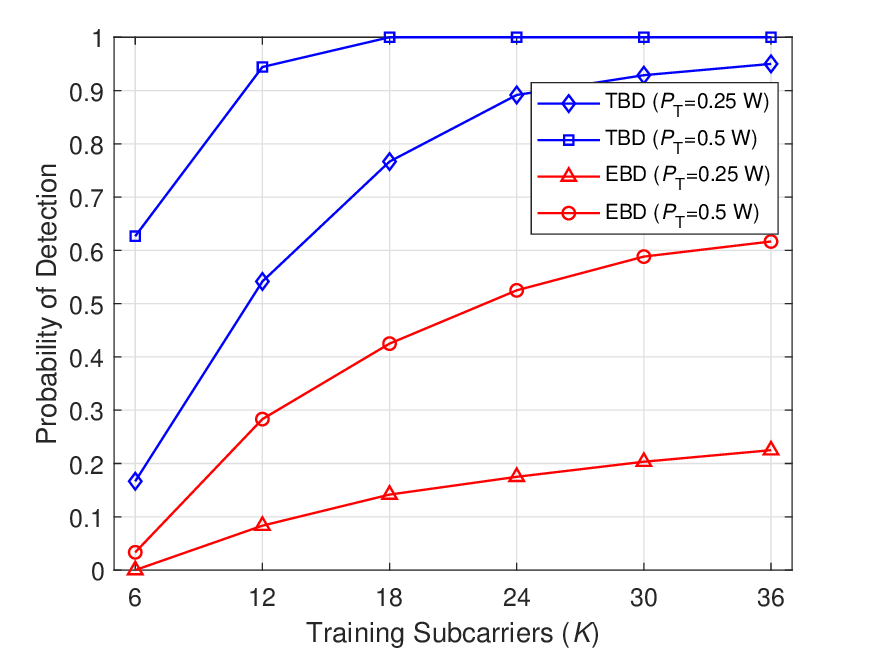}
	\caption{The effect of different power levels and number of training subcarriers ($K$) on the performance of the TBD and EBD schemes.}
	\label{result:fig2}
\end{figure}

\begingroup
\setlength{\tabcolsep}{8pt} 
\setlength{\arrayrulewidth}{0.4mm}
\renewcommand{\arraystretch}{1.4} 
\begin{table}[t!]
	\centering
	\caption{\label{table1}Parameter settings for simulation.}
	\begin{tabular}{c c c}
		\hline \textbf{Symbol} & \textbf{Parameter} & \textbf{Value} \\
		\hline $f$ & Carrier frequency & $100 \mathrm{~GHz}$ \\
		$B$ & Bandwidth & $2 \mathrm{~GHz}$ \\
		$P_{T}$ & Transmit power & {$[20 \mathrm{dBm}, 30 \mathrm{dBW}]$} \\
		${K}$ & Number of training subcarriers & $6$ \\
		$T_b$ & Solar brightness temperature & $6000 \mathrm{~K}$ \\
		$T_0$ & Ambient noise temperature in LEO & $1000 \mathrm{~K}$ \\
		$\mu$ & Regularization value for ALS & $[0.06,0.08,0.1]$ \\
		\hline
	\end{tabular}
\end{table}
\endgroup


In Fig. 2, the probability of detection ($P_D$) is presented to show the true detection at the \ac{rx} satellite while $P_M=1-P_D$ gives the probability of miss detection. It can be observed that the proposed TBD performs well by tuning the regularization parameter value to its minimum as it retrieves more fine-grained details. Furthermore, the TBD at maximum value of $\mu=0.1$ still outperforms EBD. However, the regularization parameter is very important for space debris detection where identifying rare or weak signals is crucial. The EBD analyze the presence of the space debris by comparing the energy of the received signal to a predefined threshold without classifying the number of reflected paths ($L$) for space debris detection. Therefore, it can be observed that EBD is not performing well because it is not able to detect the crucial part of the received signal which in the considered case is NLOS link due to the high path loss. Thus, it leads to a lower probability of detection than TBD as the signal of interest is weak. We remark that a higher rank of a tensor indicates a more complex channel environment with more reflected paths while a lower rank shows the LOS link having strong signal power and hence, less prone to reflection loss. Moreover, the reflection coefficient used in (8) is randomly generated from a uniform set of entries where higher value represents the space debris with highly reflective surfaces or materials and vice versa. These reflection coefficient values directly affect the pathloss that results in high or low probability of detection.

\begin{figure}[t!]
	\includegraphics[scale=0.6]{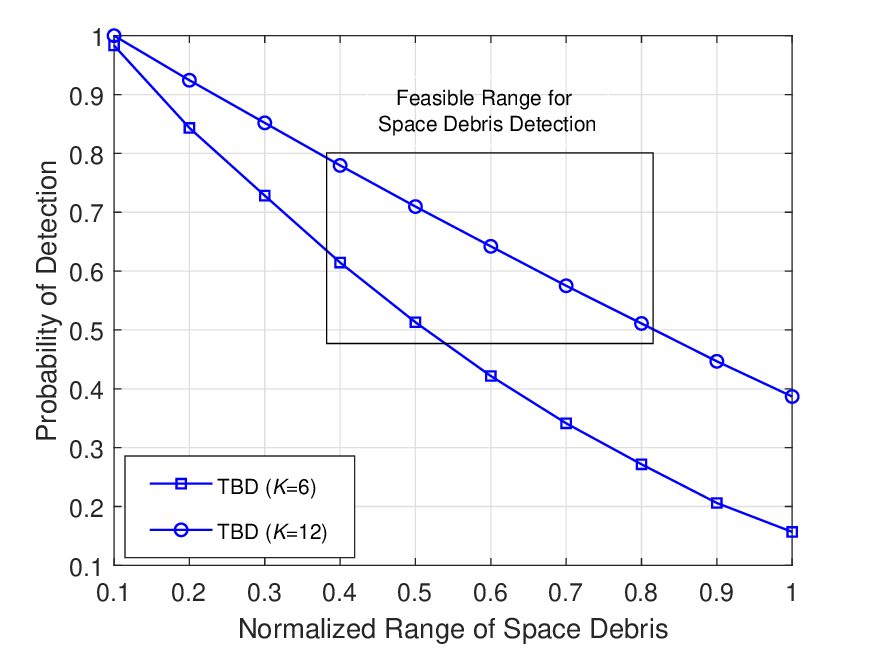}
	\caption{Feasible range analysis of the proposed TBD for space debris detection  considering $K=6,$ and $K=12$.}
	\label{result:fig3}
\end{figure}

The performance of the TBD is analyzed and compared with EBD at different power levels and training subcarriers as shown in Fig. 3. The number of training subcarriers in MIMO-OFDM system has significant impact on improving the probability of detection. The Rx satellite estimates the rank more accurately by increasing the number of training subcarriers, thus leading to more reliable signal detection. However, it is important to note that choosing the number of training subcarriers in ISAC system is a design trade-off. While more training subcarriers enhance the probability of detection, they also reduce the available bandwidth for communication.

Next, we present a feasible range to determine the safety of LEO satellites in Fig. 4. It is observed that the probability of detection of the space debris can be increased by increasing the number of training subcarriers. This is of particular interest when LEO satellites are in dense space debris regions to prevent collision at the cost of low bandwidth for communications. The optimal resource allocation for ISAC and the optimal safety range determination for space debris detection considering a large number of space debris are the topics of future work.   

\section{Conclusion}
The proliferation of space debris in LEO necessitates innovative solutions to ensure the safety and sustainability of satellite mega-constellations and future space missions. This paper proposed a ISAC technique designed to detect space debris for LEO satellites. Leveraging the sparsity of the THz channel with limited scattering, the proposed technique employed the CP tensor decomposition method, obtained using a regularized ALS algorithm, to extract the number of debris in sight. Importantly, the investigation indicated the effective utilization of NLOS links of the THz channel for space debris detection. The proposed TBD scheme outperformed conventional EBD in our numerical example. Our method thus promotes better safety and sustainability of LEO satellites and future space missions. Addressing the challenges of different types of space debris considering multi-orbit LEO satellites along with the analysis of various estimation techniques at the receiver will be part of our future works. 

\section*{Acknowledgment}
This work was supported in part by the Fonds de recherche du Québec-Nature~et~technologies (FRQNT), the Tier~1 Canada Research Chair program and the Natural Sciences and Engineering Research Council of Canada (NSERC) Discovery Grant program.

\balance
\bibliographystyle{IEEEtran}
\bibliography{references.bib}

\end{document}